\documentclass[useAMS,twocolumn,usenatbib,usegraphicx]{mn2e}
\usepackage{graphicx}

\newcommand{\go}{\mathrel{\raise.3ex\hbox{$>$}\mkern-14mu
             \lower0.6ex\hbox{$\sim$}}}
\newcommand{\lo}{\mathrel{\raise.3ex\hbox{$<$}\mkern-14mu
             \lower0.6ex\hbox{$\sim$}}}

\newcommand{\veck}{\bmath k}
\newcommand{\vecr}{\bmath r}
\newcommand{\vecB}{\bmath B}

\newcommand{\vecE}{\bmath E}
\newcommand{\vecn}{\bmath n}

\newcommand{\vecv}{\bmath v}

\newcommand{\vecb}{\bmath b}

\newcommand{\vechatr}{\hat{\bmath r}}
\newcommand{\vechatm}{\hat{\bmath m}}

\newcommand{\vechatk}{\hat{\bmath k}}

\newcommand{\vechatY}{\hat{\bmath Y}}
\newcommand{\vechatZ}{\hat{\bmath Z}}

\newcommand{\intd}{{\rm d}}

\title{Polarized Curvature Radiation in Pulsar Magnetosphere}

\author[P. F. Wang, C. Wang, and J. L. Han]
{P. F. Wang\thanks{E-mail: pfwang@nao.cas.cn}, C. Wang, and J. L. Han \\
  National Astronomical Observatories, Chinese Academy of
  Sciences.  A20 Datun Road, Chaoyang District, Beijing 100012, China \\
}

\pagerange{\pageref{firstpage}--\pageref{lastpage}}

\begin{document}

\maketitle

\label{firstpage}

\begin{abstract}
The propagation of polarized emission in pulsar magnetosphere is
investigated in this paper. The polarized waves are generated through
curvature radiation from the relativistic particles streaming along
curved magnetic field lines and co-rotating with the pulsar
magnetosphere. Within the $1/\gamma$ emission cone, the waves can be
divided into two natural wave mode components, the ordinary (O) mode
and the extraordinary (X) mode, with comparable intensities. Both
components propagate separately in magnetosphere, and are aligned
within the cone by adiabatic walking. The refraction of O-mode makes
the two components separated and incoherent. The detectable emission
at a given height and a given rotation phase consists of incoherent
X-mode and O-mode components coming from discrete emission
regions. For four particle-density models in the form of uniformity,
cone, core and patches, we calculate the intensities for each mode
numerically within the entire pulsar beam. If the co-rotation of
relativistic particles with magnetosphere is not considered, the
intensity distributions for the X-mode and O-mode components are quite
similar within the pulsar beam, which causes serious
depolarization. However, if the co-rotation of relativistic particles
is considered, the intensity distributions of the two modes are very
different, and the net polarization of out-coming emission should be
significant. Our numerical results are compared with observations, and
can naturally explain the orthogonal polarization modes of some
pulsars. Strong linear polarizations of some parts of pulsar profile
can be reproduced by curvature radiation and subsequent propagation
effect.
\end{abstract}


\section{INTRODUCTION}
Polarized pulse profiles are the key observations to understand the
physical processes within pulsar magnetosphere. Various polarization
features have been observed
\citep[e.g.][]{lm88,rr03,hmx+98,han09}. The linear polarizations are
dominating for whole or some parts of profiles, reaching $100\%$ such
as Vela and PSR B1737-30 \citep{wml93}. The position angles (PA) of
linear polarization often follow S-shaped curves, which has been well
explained by the Rotating Vector Model (RVM) \citep{rc69}. Some
pulsars exhibit two orthogonally polarized modes in the PA curves,
where a $90^\circ$ jump is accompanied by low linear polarization
\citep{scr+84,ms00}. The circular polarizations of integrated profiles
are usually weak (less than $10\%$) and diverse \citep{hmx+98}. There
exist two main types of circular polarization signature: the
antisymmetric type with a sign reversal in the mid-pulse and the
symmetric type without sign changes over the whole profile
\citep{rr90}.

To understand these diverse polarization features, there have been a
lot of theoretical works on pulsar emission processes and propagation
effects in pulsar magnetospheres. Pulsar radio emission is generally
believed to be coherent radiation from relativistic particles
streaming along the open magnetic field lines in pulsar
magnetosphere. The coherence can be caused either by the stable charge
bunches (`antenna' mechanism, e.g. \citet{bb77}) or by the
instabilities (`maser' mechanism) including the curvature plasma
instability \citep{bgi88}, two-stream instability
\citep[e.g.][]{kmm91}, the beam-plasma instability induced by the
curvature drift \citep{lmm94}, etc. Various models have been
constructed based on these two coherent manners
\citep[e.g.][]{bgi88,xlh+00,gan10}. Among these models, curvature
radiation from charge bunches serves as one of the most probable
mechanisms, whose coherent process, luminosity and spectrum have been
investigated already \citep[e.g.][]{bb76,bb77,ou80}. Its polarization
features are the most important base for theoretical
considerations. \cite{gs90} pointed out that the curvature radiation
can lead to the sign reversal feature for circular polarization of the
core component. By incorporating rotation effects, \citet{bcw91}
studied the phase lag between the centers of the PA curve and the
intensity profile. \citet{gan10} considered the detailed geometry for
curvature radiation and explained the correlation between the PA and
the sign reversal of circular polarization. Recently, \citet{wwh12}
and \citet{kg12} independently considered the co-rotation effect and
the emission geometry, and demonstrated that the circular polarization
can be of a single sign or sign reversals depending on the density
gradient of particles along the rotation phase.

In addition to the emission process, propagation effects may also have
significant influences on pulsar polarizations. The plasma within
pulsar magnetosphere has two orthogonally polarized transverse modes,
the ordinary mode (O-mode) and the extraordinary mode (X-mode)
\citep[e.g.][]{ms77,ab86,bgi93,wl07}. When emission leaves the
magnetosphere, it is coupled to these two modes. As the emitted waves
propagate outward, they first experience the ``adiabatic walking",
i.e., with their polarization following the direction of the local
magnetic field \citep{cr79}. Then, with the change of plasma
conditions, the ``wave mode coupling" occurs near the polarization
limiting radius where the mode evolution becomes non-adiabatic, which
will results in circular polarization \citep{cr79,pet06,wlh10}. As the
waves propagate further to the far magnetosphere, the ``cyclotron
absorption" takes place, which will lead to intensity absorption and
net circular polarization if the distributions of electrons and
positrons are asymmetric \citep{lm01,pet06,wlh10}. Moreover, unlike
the rectilinear propagation for the X-mode waves, the O-mode waves
will undergo refraction additionally in the near regions of their
generation \citep{ba86,lp98}, which makes the two components separated
finally.

These physical processes within pulsar magnetosphere have been studied
for the emission generation and propagation. However, the joint
researches for emission and propagation to explain observational facts
are rare. As the first try, \citet{cr79} considered the emission of the
relativistic particles subjected to parallel and perpendicular
(compare to local magnetic field) accelerations, and demonstrated the
influence of ``adiabatic walking'' on the ordering of the polarization
vectors within the $1/\gamma$ emission cones. They found that the
combination for emission in the inner magnetosphere and propagation
effects in the outer magnetosphere can naturally explain the observed
$100\%$ polarization and the sudden orthogonal mode transitions
observed for the individual pulses. However, they did not consider the
rotational effect of the relativistic particles and the refractions
for the O-mode waves. The beamed emissions within the emission cone
were improperly treated as coherent radiations. \citet{wwh12} depicted
the most detailed pictures of curvature radiation, but propagation
effects are not included yet. \citet{wlh10} and \citet{bp12} studied
the influences of the propagation effects on the mean profiles, but
the intensity and polarization of the emitted waves were assumed
rather than calculated from the emission mechanisms. Stimulated by
these researches, we focus on the general picture for the combined
effects of curvature radiation and propagation effects, and tries to
tie the theoretical explanation with polarization observations.

In this paper, we study the polarization features of the emission
generated by curvature radiation and their modifications during wave
propagation in pulsar magnetosphere through numerical calculations and
simulations. The paper is organized as follows. In Section 2, we
present the theoretical grounds of curvature radiation and propagation
effects for our calculations. Polarized emission within the $1/\gamma$
cone and the evolution of two separated natural modes in magnetosphere
are studied in Section 3. In Section 4, the intensities of both modes
within the entire pulsar beam are calculated for the density models of
uniformity, cone, core and patches. Discussion and conclusions are
given in Section 5.

\section{Theoretical basics for curvature radiation and propagation}

\subsection{Curvature radiation process}

\begin{figure}
  \centering
  \includegraphics[angle=0,width=0.48\textwidth]{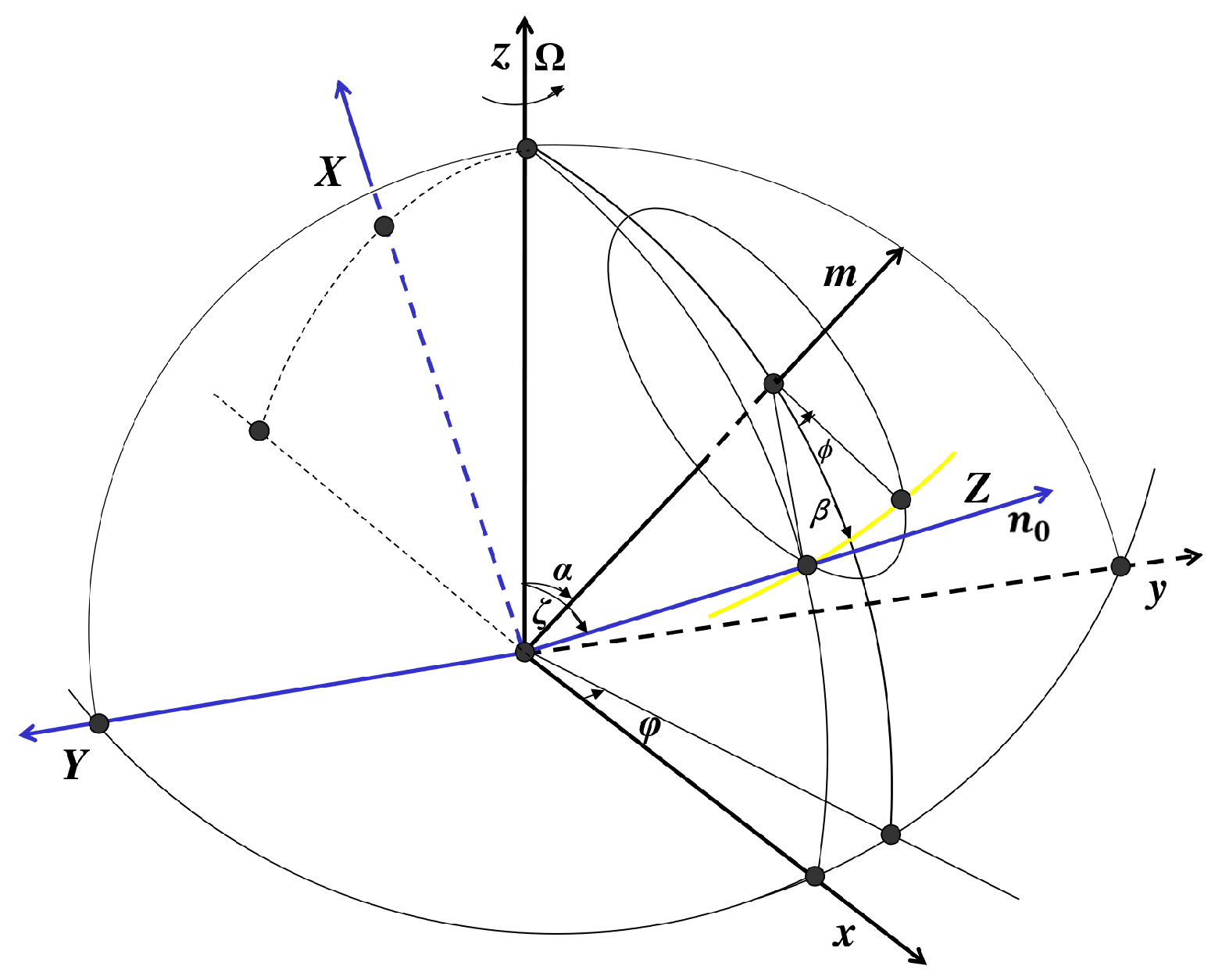}
  \caption{Coordinate systems for curvature radiation and
    propagation. The magnetic dipole moment $\bmath m$ is inclined by
    an angle of $\alpha$ with respect to the rotation axis $\bmath
    \Omega$. Sight line $\bmath n_0$ has impact angle $\beta$ and
    makes an angle $\zeta=\alpha+\beta$ with respect to $\bmath
    \Omega$. The $xyz$ coordinate system is fixed in space and used
    for the emission analysis, with $z$-axis along $\bmath \Omega$ and
    $\bmath n_0$ in $xz$-plane. The rotation phase $\varphi$ is the
    azimuthal angle of $\bmath m$ in $xyz$-frame. The $XYZ$ system is
    also a fixed frame and used for the wave propagation analysis,
    with the $Z$-axis along $\bmath n_0$, $\bmath \Omega$ in
    $XZ$-plane, and $\vechatZ\times\hat{\bmath
      \Omega}=\sin\zeta\vechatY$.}
\label{fig:geometry}
\end{figure}

In general, pulsar magnetosphere has an inclined dipole magnetic
field, which rotates uniformly in free space,
\begin{equation}
\vecB=B_{\star}(\frac{R_{\star}}{r})^3[3\vechatr(\vechatr\cdot
\vechatm)-\vechatm],
\label{eq:staticb}
\end{equation}
here $B_{\star}$ is the magnetic field on the neutron star surface,
$R_{\star}$ is the neutron star radius, $\vechatr$ is the unit vector
along $\vecr$, and $\vechatm$ represents the unit vector of the
magnetic dipole moment, as shown in Fig.~\ref{fig:geometry}. According
to \citet{rs75}, relativistic particles are generated in the vacuum
gaps above the polar caps of pulsar magnetosphere by the sparking
process and streaming out along the open magnetic field lines. Due to
the bending of the field lines, the relativistic particles will
experience perpendicular acceleration and produce curvature radiation.
The coherent bunches of the relativistic particles are simply treated
as huge point charges.

For a relativistic particle traveling with velocity $\vecv$ along an
open field line, its radiation field $\vecE(t)$ in direction
$\hat{\vecn}$ and the corresponding Fourier components $\vecE(\omega)$
have been depicted in detail by \citet{wwh12}. The energy radiated per
unit frequency per unit solid angle reads \citep{jac75}
\begin{eqnarray}
\frac{\intd^2I}{\intd\omega \intd\Omega}&=&\frac{c R^2_0}{2\pi}|\vecE(\omega)|^2 \nonumber\\
&=&\frac{e^2}{3\pi^2c}\left(\frac{\omega \rho}{c}\right)^2\left(1/\gamma^2+\theta_{nv}^2\right)^2                                                            \nonumber\\
&&[K^2_{2/3}(\xi)+\frac{\theta_{nv}^2}{1/\gamma^2+\theta_{nv}^2}K^2_{1/3}(\xi)].
\label{eq:Energy}
\end{eqnarray}
Here $R_0$ is the distance between the trajectory center and the
observer, $\rho$ is the curvature radius for the particle trajectory,
Lorentz factor $\gamma=1/\sqrt{1-v^2/c^2}$, $\theta_{nv}$ is the angle
between the emission direction $\hat{\vecn}$ and the particle velocity
$\vecv$, $\xi=\frac{\omega \rho}{3
  c}\left(1/\gamma^2+\theta_{nv}^2\right)^{3/2}$, and $K(\xi)$ are the
modified Bessel functions of the second kind.

\subsection{Propagation effects}

Immediately after the electromagnetic waves were generated through
curvature radiation from the relativistic particles, they will be
coupled to the local plasma modes to propagate outwards from the
emission regions. For the sake of simplicity, the plasma within the
magnetosphere is assumed to be cold and moves with a single velocity
$v$. The plasma is also assumed to have a density of $N_{\rm p}=\eta
N_{\rm GJ}$, where $\eta$ is the multiplicity factor, $N_{\rm
  GJ}=\Omega B/(2\pi ec)$ represents the Goldreich-Julian density
\citep{gj69}. In general, there are four wave modes within the plasma
of pulsar magnetosphere \citep{bgi93,bp12}, two transverse and two
longitudinal waves. Three out of them were commonly investigated
\citep[e.g.][]{ab86}, which are the X-mode with refraction index
$n_{\rm X}=1$, the sub-luminous O-mode with $n_{\rm O}>1$ and the
super-luminous O-mode wave with $n_{\rm O}<1$. The subluminous O-mode
(or the Alfven O-mode wave) tends to follow the field lines and
suffers serious Landau damping \citep{ba86,bgi88}, which will make
this mode invisible. Only the X-mode and superluminous O-mode
(hereafter O-mode for simplicity) can propagate outside the
magnetosphere. These two modes have different evolution behaviors for
the trajectories and polarizations.

\subsubsection{Refraction of O-mode}

The X-mode waves are always the transverse ones and propagate
rectilinearly. It is convenient to analyze the propagation process in
the fixed $XYZ$-frame, as shown in Fig.~\ref{fig:geometry}. Here, the
$Z$-axis is set along the wave vector $\bmath{k}$, e.g., the direction
for the sight line $\vecn_0$. Let the parameters with subscript ``i''
denote the quantities at the emission point. At an instant time $t$,
one photon has traveled over a distance of $s=c(t-t_{\rm i})$ along
the ray from the emission point. The position vector $\vecr$ and the
pulsar rotation phase $\varphi$ can be written as
\begin{equation}
\vecr=\vecr_{\rm i}+s\vechatZ, \nonumber
\label{eq:r}
\end{equation}
\begin{equation}
\varphi=\varphi_{\rm i}+\Omega (t-t_{\rm i})=\varphi_{\rm i}+s/r_{\rm lc},
\label{eq:phi1}
\end{equation}
here, $r_{\rm lc}$ denotes the light cylinder radius. The plasma
conditions, such as the density $N_{\rm p}$, the magnetic field
$\bmath B$, and hence the dielectric tensor $\bmath \epsilon$ vary
along the trajectory.

The O-mode wave evolves in a different manner. Refractive index of the
O-mode wave is not so close to the unity as the X-mode, especially
near the emission region with higher plasma density, which means that
it will be deflected away from the magnetic axis. When the O-mode wave
propagates outside to the low density region, the refraction effect
becomes weaker and weaker, and the trajectory gets more close to the
straight line just as the X-mode \citep{ba86,lp98,bp12}. The
trajectory for the O-mode wave can be described by the Hamilton
equations. With the dispersion relation of the mode taken into
account, the Hamilton equations become \citep{ba86},
\begin{eqnarray}
&&\frac{1}{c}\frac{\intd \vecr}{\intd t}=p\vecn-q \vecb \nonumber\\
&&\frac{1}{c}\frac{\intd \vecn}{\intd t}=q\frac{\partial(\vecb
\cdot \vecn)}{\partial \vecr}|_{\vecn}-l\alpha_{\rm p}\frac{\partial \ln
N_{\rm p}}{\partial \vecr} \label{eq:refra}
\end{eqnarray}
here, $\vecb=\vecB/|\vecB|$ is the unit vector aligned with the
magnetic field, $\vecn=c\veck/\omega$ represents the three
dimensional refractive index,
$n_\parallel=c\veck\cdot\vecb/\omega$ is the refractive index
along the magnetic field, $\alpha_{\rm p}=\omega_{\rm
p}^2/(\omega^2\gamma^3)$ with the plasma frequency $\omega_{\rm
pl}=\sqrt{4\pi N_{\rm p} e^2/m}$, $p=(1-n_{\parallel}v)^3/d$,
$q=\alpha_{\rm p}(n_\parallel-v)/d$, and
$l=(1-n_\parallel^2)(1-n_\parallel v)/(2d)$ with $d=(1-n_\parallel
v)^3-\alpha_{\rm p}n_\parallel(n_\parallel-v)$.

\begin{figure}
  \centering
  \includegraphics[angle=0,width=0.48\textwidth]{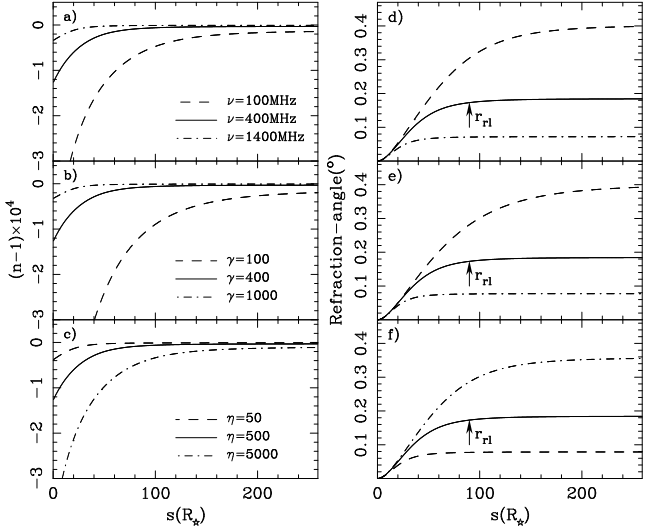}
  \caption{Evolution of the refractive index $n$ and refraction angle
    along the ray for the O-mode wave, for different observing
    frequencies $\nu$ in panels a) and d), for different Lorentz
    factors $\gamma$ in (b) and (e), for different multiplicities
    $\eta$ in (c) and (f). The refraction limiting radii, $r_{\rm rl}$,
    are indicated by the arrows in panels (d), (e) and (f) for the
    default curve (solid line in each panel). The wave is emitted from
    a height $r_{\rm i}=40R_\star$ at rotation phase $\varphi_{\rm
      i}=0^\circ$. The default parameters for the calculation are
    $\gamma=400$, $\eta=500$ and $\nu=400 \rm MHz$ for a pulsar with
    $P=1 \rm s$, $\alpha=30^\circ$ and $\beta=3^\circ$. }
  \label{fig:ref}
\end{figure}

By solving the Hamilton equations in the 3-dimensional magnetosphere,
we obtained the refractive index $n=|\vecn|$ for the O-mode wave and
the refraction angle (the angle between $\hat{\vecn}$ and initial
emission direction) along the ray, as shown in
Fig.~\ref{fig:ref}. Near the emission region, the refractive index of
the O-mode wave $n<1$, and the wave is deflected away from the
emission direction. When the photon propagates outwards, the
refraction effect can be neglected when $n\simeq1$, and then the
O-mode trajectory becomes rectilinearly, i.e., with almost the fixed
angle compared to the initial emission direction. The transition
happens at the refraction-limiting radius (RLR), $r_{\rm rl}$, where
$n_\perp^2\approx2\alpha_{\rm p}^{1/2}$ according to \citet{ba86}. It
is redefined numerically as $|(\hat{\vecn}\times\vechatk_{r_{\rm
    rl}}-\hat{\vecn}\times\vechatk_{r_{\rm
    rl}-R_\star})/(\hat{\vecn}\times\vechatk_{r_{\rm
    rl}-R_\star})|\approx10^{-3}$, as indicated in
Fig.\ref{fig:ref}. Furthermore, high frequency emission is less
refracted during its propagation, as shown in panels (a) and (b). In
addition, if the plasma within pulsar magnetosphere is thin (a small
$\eta$) and streams with a relatively larger velocity (a large Lorentz
factor $\gamma$), the O-mode wave will be less affected by the
refraction process, as shown in Fig.\ref{fig:ref} (b), (c), (e) and
(f). For the reasonable parameters chosen in Fig.\ref{fig:ref}, the
refraction-limiting radius are about 50--200$R_\ast$ and the final
refraction angles are $0.05^\circ$--$0.4^\circ$. In summary, our 3-D
refraction calculation confirms the results from the traditional 2-D
treatment, in which only the refraction within a fixed field line
plane is considered \citep[e.g.][]{ba86,bgi88,lp98,bp12}.

\begin{figure}
  \centering
  \includegraphics[angle=0,width=0.4\textwidth]{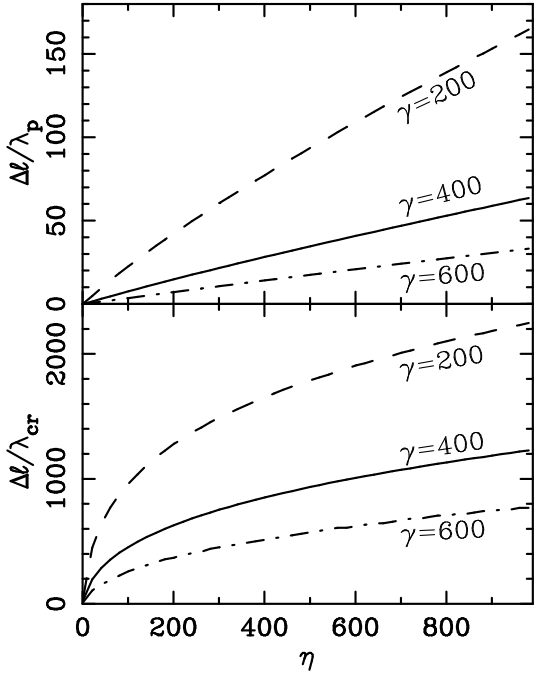}
  \caption{The separation, $\Delta l$, between the emission points of
    the O-mode and X-mode waves for various $\eta$ and $\gamma$. It is
    measured with respect to the wave lengths of the electromagnetic
    wave $\lambda_{\rm cr}$ or the plasma wave $\lambda_{\rm p}$. The
    default parameters for the calculations are the same as those in
    Fig.~\ref{fig:ref}.}
  \label{fig:mode_split}
\end{figure}

As demonstrated above, if the low frequency O-mode wave (small $\nu$)
travels through the dense (large $\eta$) and/or less relativistic
(small $\gamma$) plasma within pulsar magnetosphere, they will
experience sufficient refraction. On the contrary, the X-mode wave
travels rectilinearly regardless of the plasma conditions. Hence, for
a given sight line at a fixed rotation phase, the observed emissions
of X-mode and O-mode originate from two discrete regions separated by
$\Delta l$ at a given emission height. As well known, pulsar
radiations are coherent. To get efficient coherency, it requires the
scale of particle bunches to be much smaller than the wave length of
the emitting waves. Fig.~\ref{fig:mode_split} shows how the
separation of the emission points for the two modes (in unit of the
curvature-radiation wave-length $\lambda_{\rm cr}$ or the plasma
oscillation wave length $\lambda_{\rm p}$) varies with plasma density
$\eta$ and Lorentz factor $\gamma$. Obviously, the separation $\Delta
l$ are much larger than the wave length $\lambda_{\rm cr}$, especially
for larger $\eta$ and smaller $\gamma$. Moreover, the bunching are
induced by the plasma oscillation with a wave length $\lambda_{\rm
  p}=2\pi c\sqrt{\gamma}/\omega_{\rm p}$. The separation $\Delta l$ is
also significantly larger than $\lambda_{\rm p}$. Therefore, the
observed X-mode and O-mode waves at a fixed rotation phase are
incoherent, which means the observed radiation should be the
incoherent superposition of the two modes from two discrete emission
regions.

\subsubsection{Adiabatic walking}

\begin{figure*}
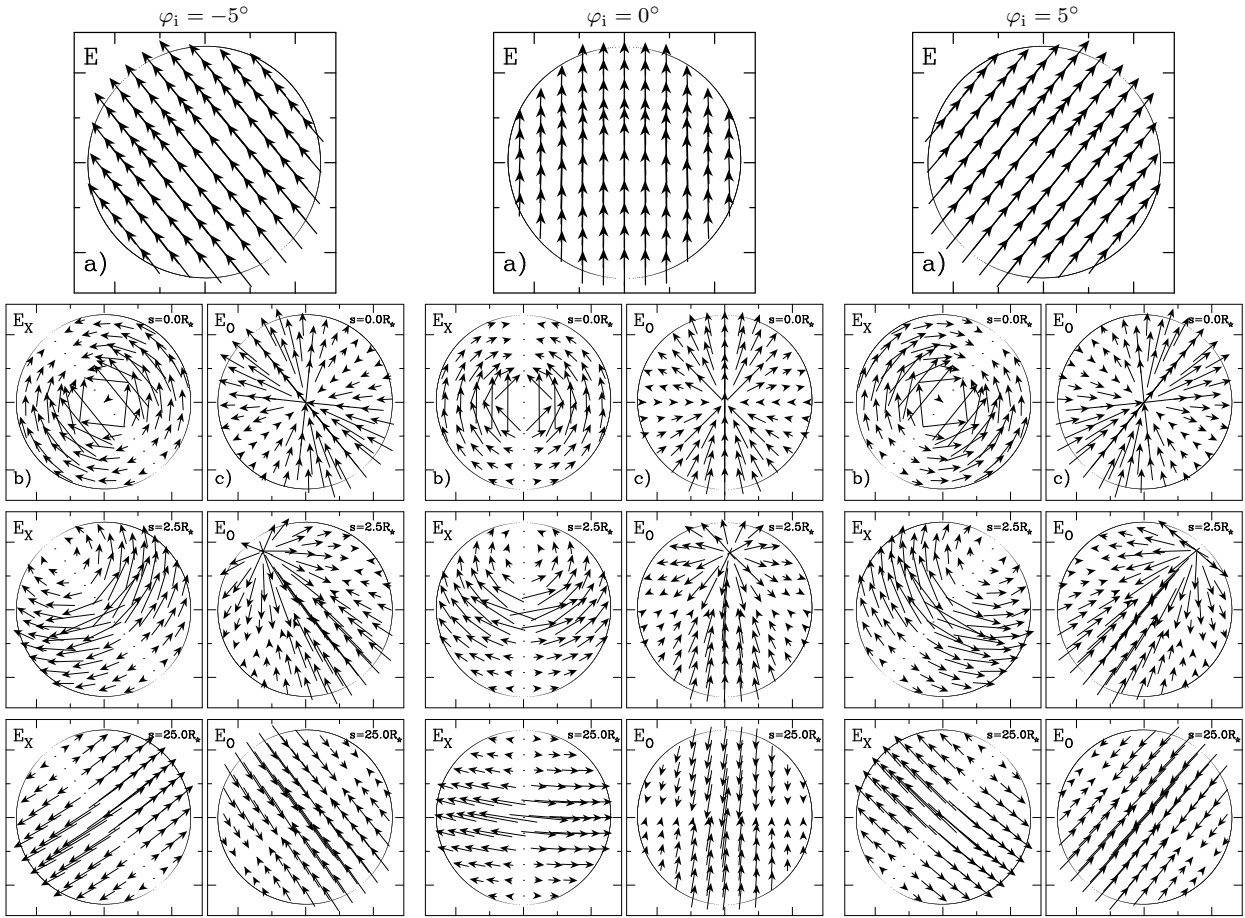

  \setlength{\tabcolsep}{0.1mm}
  \centering
  \begin{tabular}{cccccccc}
   \multicolumn{2}{c}{$\varphi_{\rm i}=-5^\circ$} & \hspace{2mm} & \multicolumn{2}{c}{ $\varphi_{\rm i}=0^\circ$} & \hspace{2mm} & \multicolumn{2}{c}{$\varphi_{\rm i}=5^\circ$}\\
    \multicolumn{2}{c}{\includegraphics[angle=0,width=0.2\textwidth]{Pattern-EP-app_E_phi-5_s0.0.ps}}&  &
    \multicolumn{2}{c}{\includegraphics[angle=0,width=0.2\textwidth]{Pattern-EP-app_E_phi0_s0.0.ps}}&  &
    \multicolumn{2}{c}{\includegraphics[angle=0,width=0.2\textwidth]{Pattern-EP-app_E_phi5_s0.0.ps}}\\
    \includegraphics[angle=0,width=0.15\textwidth]{Pattern-EP-app_EX_phi-5_s0.0.ps}&
    \includegraphics[angle=0,width=0.15\textwidth]{Pattern-EP-app_EO_phi-5_s0.0.ps}&  &
    \includegraphics[angle=0,width=0.15\textwidth]{Pattern-EP-app_EX_phi0_s0.0.ps}&
    \includegraphics[angle=0,width=0.15\textwidth]{Pattern-EP-app_EO_phi0_s0.0.ps}&  &
    \includegraphics[angle=0,width=0.15\textwidth]{Pattern-EP-app_EX_phi5_s0.0.ps}&
    \includegraphics[angle=0,width=0.15\textwidth]{Pattern-EP-app_EO_phi5_s0.0.ps}\\
    \includegraphics[angle=0,width=0.15\textwidth]{Pattern-EP-app_EX_phi-5_s2.5.ps}&
    \includegraphics[angle=0,width=0.15\textwidth]{Pattern-EP-app_EO_phi-5_s2.5.ps}&  &
    \includegraphics[angle=0,width=0.15\textwidth]{Pattern-EP-app_EX_phi0_s2.5.ps}&
    \includegraphics[angle=0,width=0.15\textwidth]{Pattern-EP-app_EO_phi0_s2.5.ps}& &
    \includegraphics[angle=0,width=0.15\textwidth]{Pattern-EP-app_EX_phi5_s2.5.ps}&
    \includegraphics[angle=0,width=0.15\textwidth]{Pattern-EP-app_EO_phi5_s2.5.ps}\\
    \includegraphics[angle=0,width=0.15\textwidth]{Pattern-EP-app_EX_phi-5_s25.0.ps}&
    \includegraphics[angle=0,width=0.15\textwidth]{Pattern-EP-app_EO_phi-5_s25.0.ps}&  &
    \includegraphics[angle=0,width=0.15\textwidth]{Pattern-EP-app_EX_phi0_s25.0.ps}&
    \includegraphics[angle=0,width=0.15\textwidth]{Pattern-EP-app_EO_phi0_s25.0.ps}& &
    \includegraphics[angle=0,width=0.15\textwidth]{Pattern-EP-app_EX_phi5_s25.0.ps}&
    \includegraphics[angle=0,width=0.15\textwidth]{Pattern-EP-app_EO_phi5_s25.0.ps}\\
  \end{tabular}
  \caption{Polarization vector distribution and propagation within the
    $1/\gamma$ emission cones without considering the co-rotation of
    particles. The initial polarization $\bmath E$ vector distributions
    are shown in panels (a) at rotation phases of $\varphi_{\rm
      i}=-5^\circ$, $0^\circ$, and $5^\circ$, and are decomposed into
    two components $\bmath E_{\rm O}$ and $\bmath E_{\rm X}$ in panels
    (b) and (c) for each rotation phase. The pointing and length of
    the arrows stand for the direction and magnitude of $\vecE$. The
    evolved distribution of X-mode and O-mode components after the
    propagation to $s=2.5 R_\star$ and $s=25 R_\star$(mainly adiabatic
    walking) are shown in the panels below for each mode. The default
    parameters for the calculation are $r_{\rm i}=40R_\star$,
    $\gamma=400$, $P=1 \rm s$, $\alpha=30^\circ$, and $\beta=3^\circ$.
  }
  \label{fig:beam-E_norot}
\end{figure*}

In addition to the ray trajectories, the polarization evolution of
X-mode and O-mode waves should be investigated. By knowing $\bmath B$,
$N_{\rm p}$, and the dielectric tensor $\bmath \epsilon$ along the
trajectory, we can use the wave equation
\begin{equation}
\nabla\times(\nabla\times \bmath E) -\frac{\omega^2}{c^2} \bmath
\epsilon \cdot \bmath E=0,
\label{eq:wave}
\end{equation}
to describe the evolution of the polarization states. By keeping the
first order terms for the transverse components of the electric field,
the wave equation (Eq.~\ref{eq:wave}) can be simplified and expressed as
the evolution equation for the eigenmode magnitudes \citep{wlh10}
\begin{equation}
i\frac{\intd}{\intd s}\left(
\begin{array}{c}
E_{\rm X}\\
E_{\rm O}\\
\end{array}
\right)
=
\left[
\begin{array}{cc}
-\Delta k/2 & i\phi'_B\\
-i\phi'_B  & \Delta k/2\\
\end{array}
\right]
\left(
\begin{array}{c}
E_{\rm X}\\
E_{\rm O}\\
\end{array}
\right),
\label{eq:eigenwave_evolv}
\end{equation}
here, $\phi'_B=d\phi_B/ds$ with the orientation of the $\veck-\vecB$
plane $\phi_B$, $\Delta k=\Delta n \omega/c=(n_{\rm X}-n_{\rm
  O})\omega/c$. To further investigate the evolution process, it is
useful to define the adiabatic parameter $\Gamma_{\rm ad}$
\citep{wlh10},
\begin{equation}
\Gamma_{\rm ad}=\left|\frac{\Delta k}{2\phi'_B}\right|.
\label{eq:gamma_ad}
\end{equation}
In the inner magnetosphere, the adiabatic condition $\Gamma_{\rm
  ad}\gg1$ is easily to be satisfied, which means the polarization
vector of the X-mode wave keeps to be orthogonal to the local
$\veck-\vecB$ plane, while that of the O-mode wave is within the
plane.  This is called the ``adiabatic walking'' process, during which
mode amplitudes keep constant along the ray while the polarization
state of each mode varies with magnetic field line. This kind of
behavior for the ``adiabatic walking'' was first investigated by
\citet{cr79} to explain the sub-pulse polarization. As the waves
travel to higher magnetosphere, $\Gamma_{\rm ad}$ gradually decreases
and finally becomes less than 1. The radius for $\Gamma_{\rm
  ad}(r_{\rm pl})=1$ is named as being the polarization limiting
radius (PLR). Near this radius, the wave-mode coupling happens, one
mode leaks to the other and vice verse, and hence circular
polarization may be generated. In the region further that $r_{\rm
  pl}$, the mode evolution becomes non-adiabatic, the polarization
states of waves are frozen and would not be affected by the plasma
(except for the possible cyclotron absorption). Generally the
polarization limiting radius, $r_{\rm pl}\sim 1000R_\ast$, is quite
far away from the emission point \citep{wlh10}. We here focus on the
influences of the initial propagation near the emission region, i.e.,
the ``adiabatic walking'' process.


\section{Emission and propagation for the polarized waves within the $1/\gamma$ cone}

A relativistic particle traveling along a curved field line will beam
its radiation around the velocity direction, forming a $1/\gamma$
emission cone. The observed radiations at a fixed rotation phase are
contributed by the particles not only at the central tangential
emission point (with velocity pointing towards the observer), but also
on the field lines within an angle of $1/\gamma$. In this section, we
will calculate the initial intensity and polarization distributions in
the $1/\gamma$ emission cone and analyze the polarization evolution of
the cone.

\subsection{Emission cone without co-rotation}

\begin{figure*}
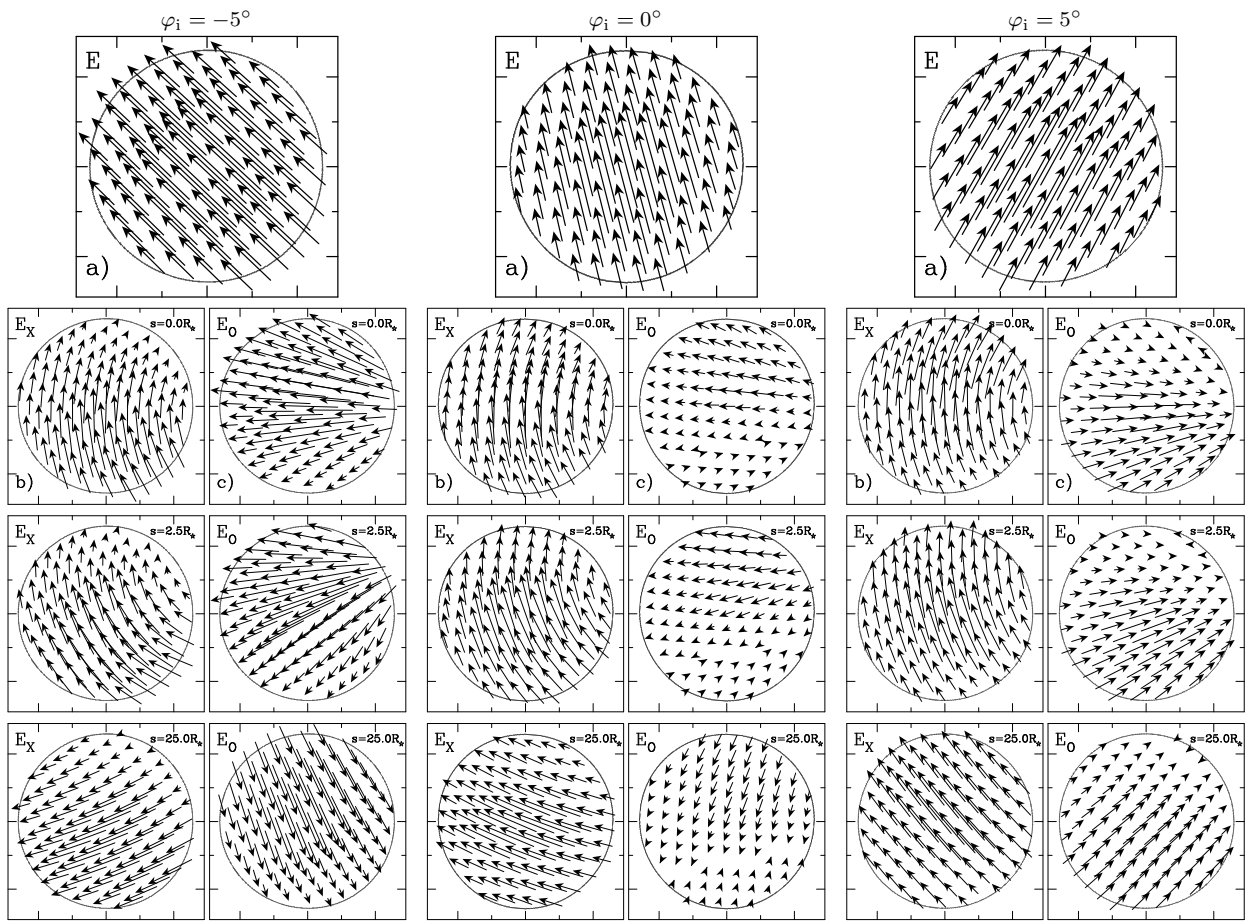

  \setlength{\tabcolsep}{0.1mm}
  \centering
  \begin{tabular}{cccccccc}
   \multicolumn{2}{c}{$\varphi_{\rm i}=-5^\circ$} & \hspace{2mm} & \multicolumn{2}{c}{ $\varphi_{\rm i}=0^\circ$} & \hspace{2mm} & \multicolumn{2}{c}{$\varphi_{\rm i}=5^\circ$}\\
    \multicolumn{2}{c}{\includegraphics[angle=0,width=0.2\textwidth]{Pattern-EP-app-E-rot_phi-5_s0.0.ps}}&  &
    \multicolumn{2}{c}{\includegraphics[angle=0,width=0.2\textwidth]{Pattern-EP-app-E-rot_phi0_s0.0.ps}}&  &
    \multicolumn{2}{c}{\includegraphics[angle=0,width=0.2\textwidth]{Pattern-EP-app-E-rot_phi5_s0.0.ps}}\\
    \includegraphics[angle=0,width=0.15\textwidth]{Pattern-EP-app-EX-rot_phi-5_s0.0.ps}&
    \includegraphics[angle=0,width=0.15\textwidth]{Pattern-EP-app-EO-rot_phi-5_s0.0.ps}&  &
    \includegraphics[angle=0,width=0.15\textwidth]{Pattern-EP-app-EX-rot_phi0_s0.0.ps}&
    \includegraphics[angle=0,width=0.15\textwidth]{Pattern-EP-app-EO-rot_phi0_s0.0.ps}&  &
    \includegraphics[angle=0,width=0.15\textwidth]{Pattern-EP-app-EX-rot_phi5_s0.0.ps}&
    \includegraphics[angle=0,width=0.15\textwidth]{Pattern-EP-app-EO-rot_phi5_s0.0.ps}\\
    \includegraphics[angle=0,width=0.15\textwidth]{Pattern-EP-app-EX-rot_phi-5_s2.5.ps}&
    \includegraphics[angle=0,width=0.15\textwidth]{Pattern-EP-app-EO-rot_phi-5_s2.5.ps}&  &
    \includegraphics[angle=0,width=0.15\textwidth]{Pattern-EP-app-EX-rot_phi0_s2.5.ps}&
    \includegraphics[angle=0,width=0.15\textwidth]{Pattern-EP-app-EO-rot_phi0_s2.5.ps}& &
    \includegraphics[angle=0,width=0.15\textwidth]{Pattern-EP-app-EX-rot_phi5_s2.5.ps}&
    \includegraphics[angle=0,width=0.15\textwidth]{Pattern-EP-app-EO-rot_phi5_s2.5.ps}\\
    \includegraphics[angle=0,width=0.15\textwidth]{Pattern-EP-app-EX-rot_phi-5_s25.0.ps}&
    \includegraphics[angle=0,width=0.15\textwidth]{Pattern-EP-app-EO-rot_phi-5_s25.0.ps}&  &
    \includegraphics[angle=0,width=0.15\textwidth]{Pattern-EP-app-EX-rot_phi0_s25.0.ps}&
    \includegraphics[angle=0,width=0.15\textwidth]{Pattern-EP-app-EO-rot_phi0_s25.0.ps}& &
    \includegraphics[angle=0,width=0.15\textwidth]{Pattern-EP-app-EX-rot_phi5_s25.0.ps}&
    \includegraphics[angle=0,width=0.15\textwidth]{Pattern-EP-app-EO-rot_phi5_s25.0.ps}\\
  \end{tabular}
  \caption{Same as Fig.~\ref{fig:beam-E_norot} except that the
    co-rotation of particles is considered. }
  \label{fig:beam-E_rot}
\end{figure*}

Consider the curvature radiation generated by the relativistic
particles without rotation, i.e., $v_r=0$. In this case, the
acceleration vector of particles locates inside the curved magnetic
field line plane. Fig.~\ref{fig:beam-E_norot}(a) shows the
distribution of the electric field vectors in the $1/\gamma$ emission
cone at a given rotation phase of $\varphi_{\rm i}=-5^\circ$,
$0^\circ$, and $5^\circ$. Due to nearly the same accelerations within
the cone, $\bmath E$ are pointing towards almost the same direction at
a given rotation phase. The electric fields $\bmath E$ could be
decomposed into O-mode components, $\bmath E_{\rm O}$, and X-mode
components, $\bmath E_{\rm X}$, as shown in the panels of
Fig.~\ref{fig:beam-E_norot} (b) and (c). The magnitudes for the O-mode
components $\bmath E_{\rm O}$ reach their maximum only in the central
magnetic field line plane of each emission cone, where $\bmath E_{\rm
  X}$ always approach zero. However, in the central plane orthogonal
to the central magnetic field line plane, $\bmath E_{\rm X}$
components reach the maximum, while the $\bmath E_{\rm O}$ components
go to almost zero.

Once the polarized waves are generated, they leave the emission
region. The trajectory and polarization evolution of a single wave
have been described in the previous analysis. For the emitted waves
within the $1/\gamma$ cone, their polarizations are greatly affected
by the ``adiabatic walking'' during the initial propagations. The
polarization evolutions of the emitted waves are shown in panels below
Fig.~\ref{fig:beam-E_norot}(b) for X-mode components and
Fig.~\ref{fig:beam-E_norot}(c) for O-mode components. When the waves
are generated at $s=0R_\star$, the polarization vectors for the X-mode
components distribute symmetrically around the central tangential
emission point (with $\veck=\vecb$). Due to the bending of the field
lines, the tangential point will gradually evolve out of the
$1/\gamma$ cone as the waves propagate outwards. When the waves
propagate to about the distance $s=25.0R_\star$, their polarization
vectors are tending to be aligned. O-mode components experience the
similar adiabatic walking as X-mode, as shown in the panels below
Fig.~\ref{fig:beam-E_norot}(c). Finally, the O-mode electric field
components are ordered to point towards the central plane (at
$s=25.0R_\star$), which is orthogonal to that for the X-mode
waves. The difference is that the trajectories of O-mode components
are refracted during the initial propagations, which make the emission
cones of X-mode and O-mode separated. Note that the final total
intensity of the cone are almost the same for X-mode and O-mode
components, which would cause strong depolarization for the final
polarization profiles. We conclude therefore that without co-rotation
the polarized emission generated in the inner magnetosphere can be
completely depolarized.


\subsection{Emission cone with co-rotation}

\begin{figure*}
  \centering
  \includegraphics[angle=0,width=0.65\textwidth]{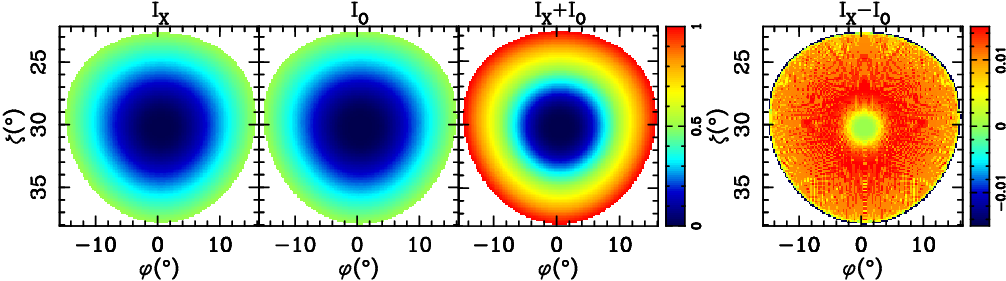} \\
  (a) Uniform density.\\
  \includegraphics[angle=0,width=0.65\textwidth]{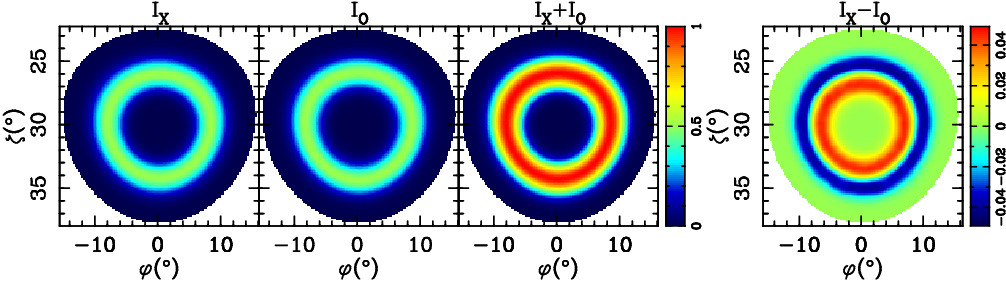}\\
  (b) Conal density.\\
  \includegraphics[angle=0,width=0.65\textwidth]{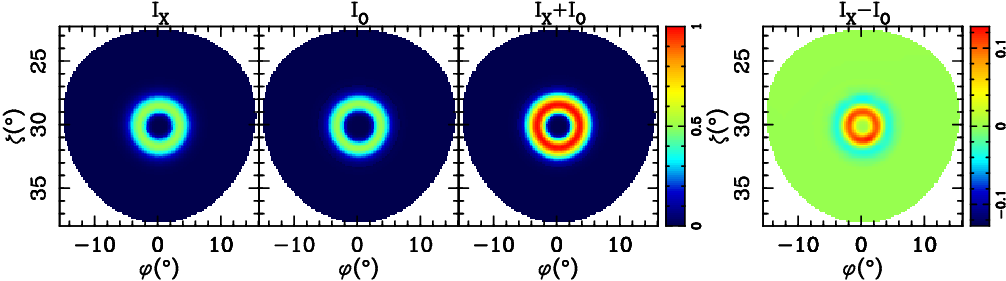} \\
  (c) Core density.\\
  \includegraphics[angle=0,width=0.65\textwidth]{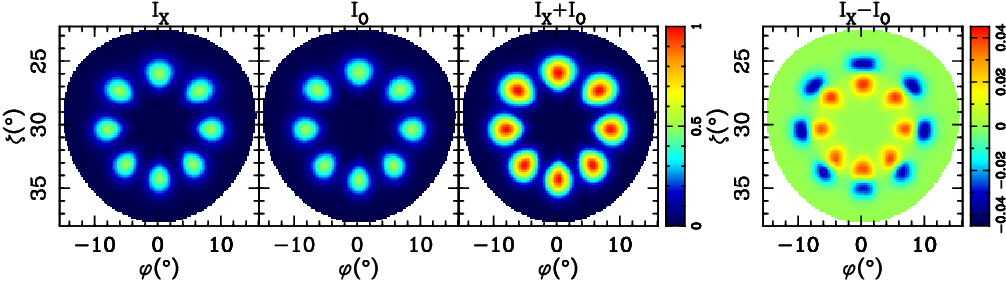}\\
  (d) Patch density.\\
  \caption{Pulsar polarization emission beams of the X-mode components
    ($I_{\rm X}$), the O-mode components ($I_{\rm O}$), the total
    intensity ($I_{\rm X}+I_{\rm O}$) and their difference ($I_{\rm
      X}-I_{\rm O}$, or the net linear polarization) for the four
    density models in the form of uniformity, cone, core and
    patches. Here, the rotations of the relativistic particles are not
    considered in the calculations. The density cone is located at
    $\vartheta_p=0.5$ and with a width of $\sigma_\vartheta=0.12$. The
    eight density patches in panel (d) are located at
    $\phi_p=0^\circ$, $\pm45^\circ$, $\pm90^\circ$, $\pm135^\circ$ and
    $180^\circ$, with $\vartheta_p=0.5$ and
    $\sigma_\vartheta=0.12$. The emission is generated at $r_{\rm
      i}=40R_\star$ with the parameters $\alpha=30^\circ$,
    $\gamma=400$, $\eta=500$, and $\nu=400 \rm MHz$. }
  \label{fig:beam_no-rot}
%
  \setlength{\tabcolsep}{0.9mm}
  \centering
  \begin{tabular}{cccc}
    \includegraphics[angle=0,width=0.24\textwidth]{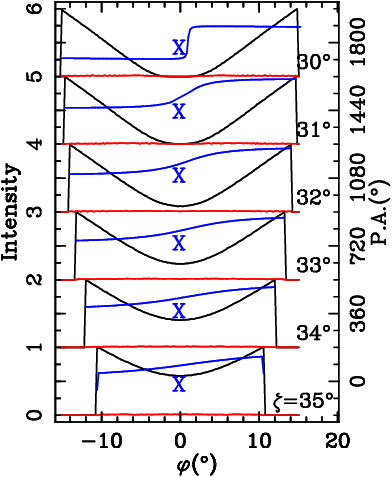} &
    \includegraphics[angle=0,width=0.24\textwidth]{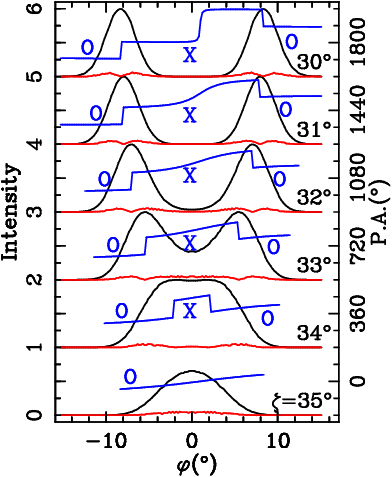}  &
    \includegraphics[angle=0,width=0.24\textwidth]{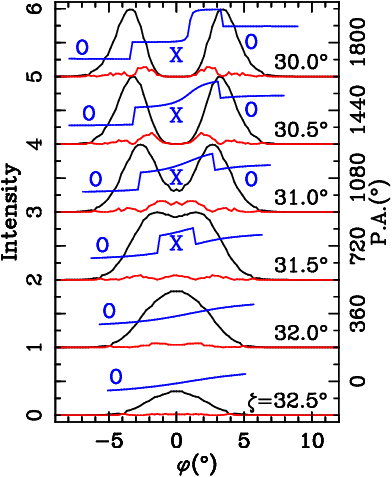}  &
    \includegraphics[angle=0,width=0.24\textwidth]{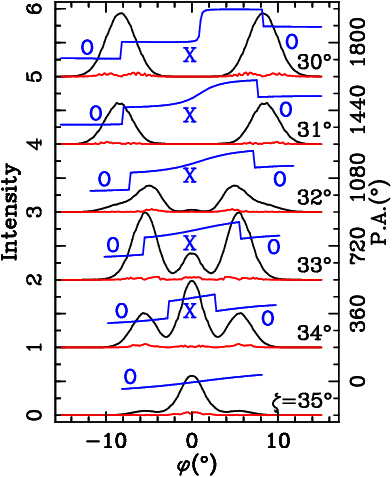} \\
    (a) Uniform density. & (b) Conal density. &  (c) Core density. &   (d) Patch density.\\
  \end{tabular}
  \caption{Pulsar polarization profiles for several fixed sight lines
    (different $\zeta$) in Fig.~\ref{fig:beam_no-rot}. The black solid
    lines denote the total intensity, $I_{\rm X}+I_{\rm O}$, the red
    lines are for the linear polarization intensities, $|I_{\rm
      X}-I_{\rm O}|$, and the blue lines represent the PA of the
    linear polarizations. The polarization modes (X or O) are marked
    near the PA curves. Due to the quasi symmetry of the beams, the
    profiles of $\zeta<30^\circ$ are similar as those of
    $\zeta>30^\circ$ except with the decreasing PA curves.}
  \label{fig:profiles-norot}
\end{figure*}

The relativistic particles in pulsar magnetosphere not only stream
along the magnetic field lines, but also co-rotate with the
magnetosphere, i.e., with $v_r\neq0$. In this case, the curvature
radiation should be different because of the additional co-rotation
velocity and acceleration. The emission process with $v_r\neq0$ was
first investigated by \citet{bcw91} to calculate the emissions from
the tangential emission points, i.e., the beam centers. \citet{wwh12}
and \citet{kg12} further analyzed the analysis to the emissions within
the entire $1/\gamma$ cone. Fig.~\ref{fig:beam-E_rot}(a) demonstrates
the co-rotation-considered curvature emission within the $1/\gamma$
cone, showing the distribution of electric field vectors $\bmath E$,
their O-mode and X-mode components ($\bmath E_{\rm O}$ and $\bmath
E_{\rm X}$). In general, the electric fields $\bmath E$ within each
cone are pointing towards almost the same direction at a given
rotation phase. However, due to the rotation induced aberration, the
emissions are shifted to the early rotation phases. It leads that the
symmetry breaks for the patterns of $\bmath E$ about the central
rotation phase $\varphi_{\rm i}=0^\circ$. Furthermore, the
polarization patterns for $\bmath E_{\rm X}$ and $\bmath E_{\rm O}$
also changed, as shown in the panels of Fig.~\ref{fig:beam-E_rot}(b)
and (c). Compared with the patterns without rotation
(Fig.~\ref{fig:beam-E_norot}), the central tangential point (where
$\vecv \parallel \vecB$) lies out of the $1/\gamma$ cone even at the
emission position due to the rotation induced aberration. After the
adiabatic walking, the two components are ordered to almost the same
direction and orthogonal with each other (see the panels below
Fig.~\ref{fig:beam-E_rot}(b) and (c)). These X-mode and O-mode
patterns just keep the polarization features of the left part of the
corresponding emission cones without rotation. The final total
intensity of the cone are quite different for the X-mode and O-mode
components due to the co-rotation induced distortion to the emission
cone patterns. Thus the observed mixed profiles would show strong net
linear polarization, which is very different from the case without
co-rotation.

\section{Pulsar emission beams of two orthogonal modes}

In addition to the analysis of emission and propagation for a single
photon and the waves within the $1/\gamma$ emission cone, we now
investigate the entire pulsar emission beam to get the polarization
pulsar profiles. For a given point within pulsar beam, the observed
radiation is the integration of all the emissions within the
$1/\gamma$ cone around the tangential direction. Here we consider the
emission from a fixed direction with a sight line angle $\zeta$ and a
rotation phase $\varphi$. Since the typical size of the emission cone
is much larger than the size of the coherent particle bunch,
incoherent addition of the radiations within the cone is
reasonable. The total intensity for the integrated emission reads,
\begin{eqnarray}
I_r(\omega) = \frac{2\pi}{c R_0^2}\int^{\theta_0+\delta\theta}_{\theta_0-\delta\theta}
\int^{\phi_0+\delta\phi}_{\phi_0-\delta\phi}
\frac{d^2I}{d\omega d\Omega} N(r,\theta,\phi)~r^2 \sin \theta \intd \theta \intd \phi.
\label{eq:intensity}
\end{eqnarray}
Here, ($\theta_0$, $\phi_0$) is the central direction of the emission
cone, while $\delta\theta$ and $\delta\phi$ denote the boundary for
the cone, $N(r,\theta,\phi)$ represents the density distribution for
the plasma within pulsar magnetosphere. We can get pulsar beam
patterns by calculating the emission intensity in every direction.

\subsection{Emission beams without co-rotation}

\begin{figure*}
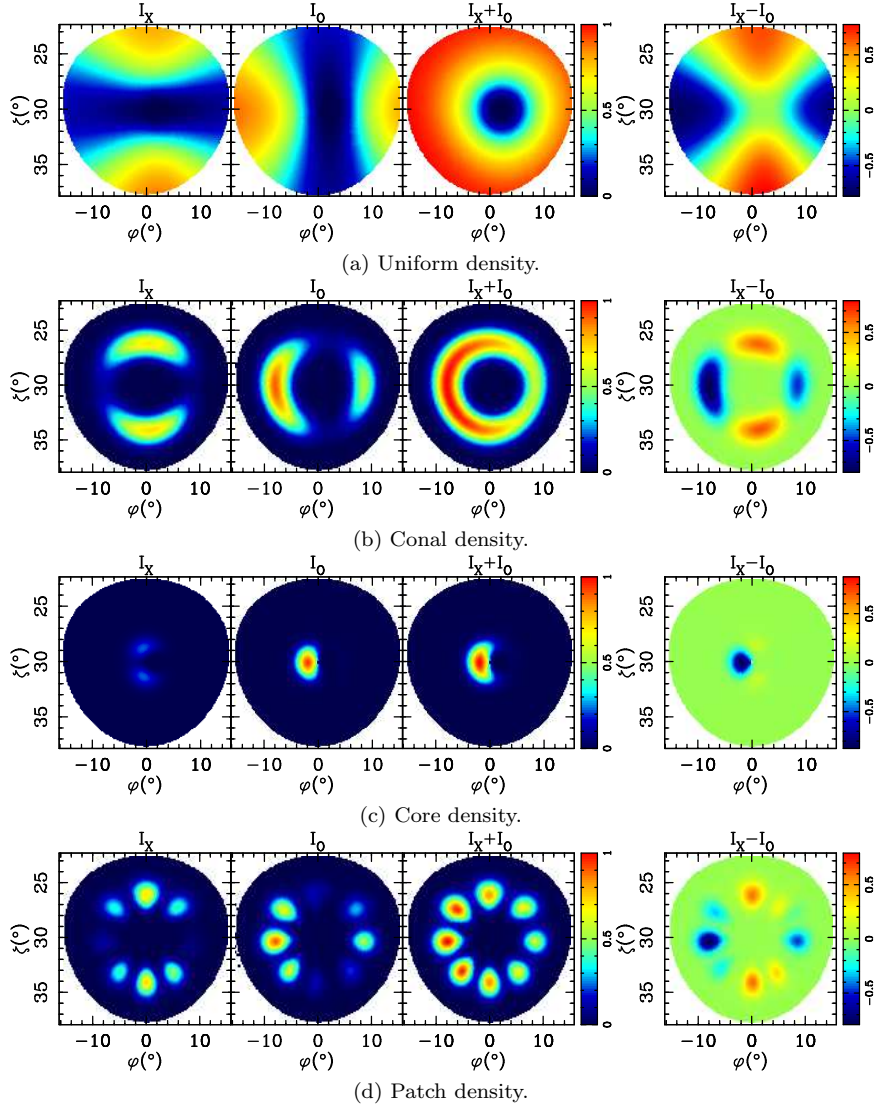

  \centering
  \includegraphics[angle=0,width=0.65\textwidth]{refuniform_rot.ps} \\
  (a) Uniform density.\\
  \includegraphics[angle=0,width=0.65\textwidth]{refcone_rot.ps}\\
  (b) Conal density.\\
  \includegraphics[angle=0,width=0.65\textwidth]{refcore_rot.ps} \\
  (c) Core density.\\
  \includegraphics[angle=0,width=0.65\textwidth]{refpatch_rot.ps}\\
  (d) Patch density.\\
  \caption{Same as Fig.~\ref{fig:beam_no-rot} except that the
    co-rotation of the relativistic particles is considered.}
  \label{fig:beam_rot}
\end{figure*}

\begin{figure*}
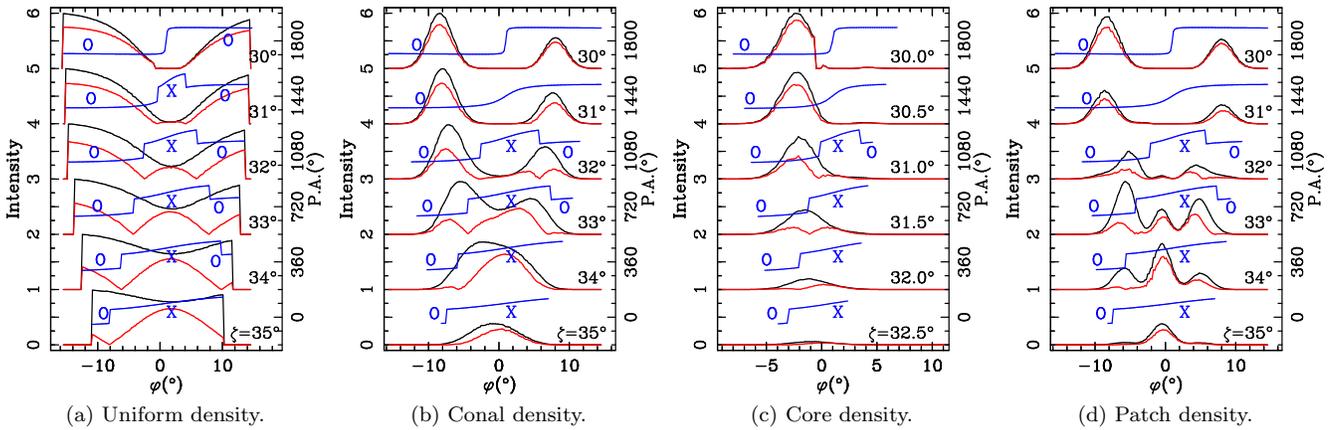

  \setlength{\tabcolsep}{0.9mm}
  \centering
  \begin{tabular}{cccc}
    \includegraphics[angle=0,width=0.24\textwidth]{Profile_rot_uniform.ps} &
    \includegraphics[angle=0,width=0.24\textwidth]{Profile_rot_cone.ps} &
    \includegraphics[angle=0,width=0.24\textwidth]{Profile_rot_core.ps} &
    \includegraphics[angle=0,width=0.24\textwidth]{Profile_rot_patch.ps} \\ 
    (a) Uniform density. & (b) Conal density. & (c) Core density.  & (d) Patch density.\\
  \end{tabular}
  \caption{Same as Fig.~\ref{fig:profiles-norot} for selected sight
    lines ($\zeta=30^o-35^o$) in Fig.~\ref{fig:beam_rot}. }
  \label{fig:profiles}
\end{figure*}

Using equation~(\ref{eq:intensity}), we can calculate the polarization
emissions from the whole open field line region without considering
the co-rotation of the relativistic particles.
Fig.~\ref{fig:beam_no-rot} shows the calculated intensity and
polarization beams for the four density models of the out-flowing
particles in the form of uniformity, cone, core and patches. For the
emission at a given point of pulsar beam, the X-mode and O-mode
components ($I_{\rm X}$ and $I_{\rm O}$) have a similar intensity
due to the quasi symmetric geometry of the magnetic field geometry
within the $1/\gamma$ emission cone. Furthermore, as mentioned
previously, the X-mode waves propagate rectilinearly, while the O-mode
waves suffer refraction. Hence, the patterns for $I_{\rm X}$ and
$I_{\rm O}$ are slightly different due to the refraction. Generally
speaking, the O-mode image is larger than that for the X-mode. For the
moderate plasma parameters used in the calculations, the refraction
effect is weak and the refraction induced differences are hard to see
directly from the patterns for $I_{\rm X}$ and $I_{\rm O}$. However,
it becomes clear from the intensity differences $I_{\rm X}-I_{\rm O}$,
which are shown in the right panels. They demonstrate the outward
shift of $I_{\rm O}$ with respect to $I_{\rm X}$ due to refraction. If
there is no refraction, $I_{\rm X}-I_{\rm O}$ will almost be zero for
the entire beam.

For the uniform density model with $N(r,\theta,\phi)=N_{\rm p}$ in the
open field line region at a given emission height, the emission
patterns are shown in Fig.~\ref{fig:beam_no-rot}(a). As we see, $I_{\rm
  X}$ and $I_{\rm O}$ are stronger in the outer parts of pulsar beam
where the curvature radii for the particle trajectories are smaller
and curvature radiation should be stronger naturally. In addition, due
to the refraction induced outward bending of the O-mode trajectories,
the intensity differences, $I_{\rm X}-I_{\rm O}$, keep to be positive
for almost all parts of the beam. Their largest difference is $1.5\%$
in Fig.~\ref{fig:beam_no-rot} for the typical parameters. Hence, the
emergent radiation suffers serious depolarization due to the
incoherent addition of the two orthogonal modes, as shown in
Fig.~\ref{fig:profiles-norot}(a). The net polarization within the entire
beam is of X-mode.

The out-flowing particles can have a variable distribution in the
polar and azimuthal directions in the magnetic axis frame, i.e.,
$N(r,\theta,\phi)=N_{\rm p}f(\theta)g(\phi)$. The density models for
the particles in the form of cone, core and patches have been widely
used in the studies of pulsar emissions
\citep[e.g.][]{wwh12,bp12}. The conal shaped density model can be
defined on the neutron star surface and extended to the high
magnetosphere, described as,
\begin{eqnarray}
f(\theta)&=&f_0\exp[-\frac{(\vartheta-\vartheta_p)^2}{2\sigma_\vartheta^2}],\nonumber\\
g(\phi)&=&1.
\label{eq:conal}
\end{eqnarray}
Here, $\vartheta=\theta_c/\theta_{c,max}$, $\theta_c$ is the polar
angle of a field line footed on the neutron star surface,
$\theta_{c,max}=\sin^{-1}(\sqrt{R_\star/r_{\rm e,lof}})$ represents
the polar angle maximum, i.e., the polar angle for the last open field
line footed on the neutron star surface. For the conal density model,
the intensities of the emitted wave are shown in
Fig.~\ref{fig:beam_no-rot}(b). Obviously the emissions are peaked at the
density cone for both X-mode and O-mode components.  The intensity
differences, $I_{\rm X}-I_{\rm O}$, exhibit sense reversals due to the
refraction of the O-mode component. Therefore, the net polarization
exhibits mode jumps twice when a sight line cuts across the beam, as
shown in Fig.~\ref{fig:profiles-norot}(b) for $\zeta=30^\circ$,
$31^\circ$, $32^\circ$, $33^\circ$, and $34^\circ$. Note that the
X-mode dominates the central parts of the polarization profiles while
O-mode dominates the two profile wings.  Only for a sight line cutting
across the edge part of the cone, for example, $\zeta=35^\circ$, the
pure O-mode polarization can be detected.

The core density model can also be described by Eq.~\ref{eq:conal}
except with $\vartheta_p=0$, the emission patterns and pulse profiles
for which are shown in Fig.~\ref{fig:beam_no-rot}(c) and
Fig.~\ref{fig:profiles-norot}(c). The intensity distribution of both
X-mode and O-mode components are similar as those for the conal
density model. The intensity near the beam center is still very
weak. Because the curvature radiation intensity is still negligible
due to the very large curvature radii (infinity for the central point
which stands for the magnetic axis), although the density for the
particles is peaked at the beam center.

The patch density model can also be described by Eq.~\ref{eq:conal},
except that $g(\phi)=g_0\exp[-(\phi-\phi_p)^2/2\sigma_\phi^2]$. Here,
$\phi_p$ represents the position for the density patch in the magnetic
azimuth direction and $\sigma_\phi$ is the characteristic width of the
patch. For the eight density patches located at different azimuthal
directions around the magnetic axis, their emission patterns and pulse
profiles are shown in Fig.~\ref{fig:beam_no-rot}(d) and
Fig.~\ref{fig:profiles-norot}(d). The X-mode components are stronger
in the central regions near the magnetic axis, while the O-mode
components are stronger in the outer parts of pulsar beam, as shown by
the pattern for $I_{\rm X}-I_{\rm O}$. The refraction behaves
similarly as in the density cone.

To summarize, the refraction will bend the O-mode components to the
outer parts of the open field line region. It leads the intensity
differences for the X-mode and O-mode components at a given position
and hence causes serious depolarization. Generally the X-mode
dominates the central parts of the beam while the O-mode dominates the
outer parts.

\subsection{Emission beams with co-rotation}

When the relativistic particles stream out in pulsar magnetosphere,
the influences of co-rotation on the emissions can not be
ignored. Here, we calculate the emissions of particles in the whole
open field line region for the density models of uniformity, cone,
core and patches. The results are shown in Fig.~\ref{fig:beam_rot} and
Fig.~\ref{fig:profiles}.

For the uniform density model, the total intensities, $I_{\rm
  X}+I_{\rm O}$, are stronger at the edge parts of the beam, similar
as the case without rotation. However, due to the rotation-induced
bending of the particle trajectories, the pattern for $I_{\rm
  X}+I_{\rm O}$ is not symmetric around the beam center, the leading
part is stronger than the trailing part. The X-mode components,
$I_{\rm X}$, are stronger at the two sides of the beam in the $\zeta$
direction, while the O-mode components are stronger at the two sides
of the beam in the $\varphi$ direction. Hence, the intensity
differences, $I_{\rm X}-I_{\rm O}$, show quadruple features. When
sight lines of different angles of $\zeta$ cut across the beam,
various profiles are shown in Fig.~\ref{fig:profiles}(a). For the
profiles detected by the sight lines except $\zeta=30^\circ$, the
X-mode component appears at the central rotation phase, while the
O-mode component emerges at two sides of the profiles.

For the conal density model, the emissions are mainly detected near
the density cone (see Fig.~\ref{fig:beam_rot}b). When the sight lines
cut across the central part of the beam (e.g. $\zeta=30^\circ$ and
$31^\circ$), the two intensity peaks are dominated by O-mode
components with a high linear polarization (see
Fig.~\ref{fig:profiles}b). However, when the sight lines cut across
the edge parts of the beam, e.g. $\zeta=34^\circ$ or $35^\circ$,
X-mode dominates the single intensity peak. For a modest $\zeta$,
e.g. $\zeta=32^\circ$ and $33^\circ$, the polarization profile shows
changes of modes as ``O~X~O'' in phases, with the X-mode dominating
the central part and the O-mode the two profile wings. This ``O~X~O''
structure of the mean profile was also predicted by \citet{bp12}, and
also comparable to observations \citep{bip13,w14}.

For the core density model, the emission patterns and pulse profiles
are shown in Fig.~\ref{fig:beam_rot}(c) and
Fig.~\ref{fig:profiles}(c). Unlike the emissions from the particles
without rotation, the relativistic particles traveling along the
central magnetic field lines also produce considerable radiation. The
emission direction is bent towards the rotation direction with the
dominating O-mode. When sight lines cut across the beam, the X-mode
emission is detected in the central part, and the O-mode emissions
from the two sides.

For eight density patches located on pulsar polar cap, the emission
beam patterns and pulse profiles are shown in Fig.~\ref{fig:beam_rot}(d)
and Fig.~\ref{fig:profiles}(d). The distributions of $I_{\rm X}$,
$I_{\rm O}$, $I_{\rm X}+I_{\rm O}$, and $I_{\rm X}-I_{\rm O}$ for the
eight density patches are similar as those discrete parts of the
density cone.

In summary, the rotation has significant influences on pulsar emission
beam and polarization. It causes the emissions of the leading
components stronger than the trailing ones. Though PA curves are quite
similar to these without rotation, the pulse profiles have significant
polarization in some parts, different from the case without rotation.

\section{Discussions and Conclusions}

In this paper, we have investigated the emission and the propagation
of a single photon, the waves within the $1/\gamma$ emission cone, and
the emission within the entire open field line region. The polarized
waves are generated through curvature radiation from relativistic
particles traveling along the curved magnetic field lines and
co-rotating with a pulsar. Once the polarized waves are generated,
they will be coupled to the local plasma modes (X-mode and/or O-mode)
to propagate outwards in pulsar magnetosphere. The X-mode component
propagates outside in a straight line, while the O-mode component
suffers the refraction as described by the Hamilton equations and its
trajectory bends towards outside, which causes the two components
separated from each other. Both X-mode and O-mode components should
experience the ``adiabatic walking'' with the polarization vectors
following the orientation of magnetic fields along the ray. We
calculated the emission and propagation of the polarized waves within
$1/\gamma$ around the velocity direction of the relativistic streaming
particles for a given rotation phase and emission height. Furthermore,
for different density models of particles in the form of the
uniformity, cone, core and patches within the open field line region,
we calculated the intensities for the X-mode and O-mode components,
$I_{\rm X}$ and $I_{\rm O}$, as well as the total intensity, $I_{\rm
  X}+I_{\rm O}$, and the linear polarization, $|I_{\rm X}-I_{\rm O}|$,
within the emission beam (Fig.~\ref{fig:beam_no-rot} and
\ref{fig:beam_rot}). When sight lines cut across the beam, pulse
profiles show various polarizations (Fig.~\ref{fig:profiles}). We
draw the following conclusions.

\begin{enumerate}
\item Both the X-mode and O-mode wave components can be produced by
  the curvature radiation process. Their intensities vary a lot for
  different emission points, but always could be comparable for any
  emission cone.
\item The refraction will bend the O-mode emissions towards the outer
  part of pulsar beam, which is serious for low frequency wave (low
  $\nu$) if the plasma within pulsar magnetosphere has a high density
  ($\eta$) and a small velocity ($\gamma$). Thus, the X-mode and
  O-mode components of initial emission are separated in pulsar
  magnetosphere, the observed polarization intensity at a given
  rotation phase should be the incoherent mixture of the X-mode and
  O-mode emission cones from discrete emission positions. {\it The
  orthogonal mode jumps happen naturally due to the change of the
  dominance of the two modes.}
\item The co-rotation of particles with magnetosphere has significant
  influence on pulsar polarized emissions. If the co-rotation is not
  considered, the observed magnitude of X-mode and O-modes within the
  emission beam should be comparable, which means a serious
  depolarization for pulsar profiles as shown in
  Fig.~\ref{fig:beam_no-rot} and \ref{fig:profiles-norot}. However, if
  the co-rotation is incorporated, the magnitude distributions for the
  X-mode and O-mode components appear in quite different regions of
  the pulsar emission beam (see Fig.~\ref{fig:beam_rot} and
  \ref{fig:profiles}), which means the final profiles could have
  significant polarization with different modes.
\item Since the refraction bend the O-mode emissions towards the outer
  part of pulsar beam, the final polarization profiles usually show
  the ``O~X~O'' modes in phases, with the X-mode dominating the
  central part and the O-mode at the two wings, which agrees with
  predictions by \citet{bp12} and observations shown by
  \citet{bp11,bip13} and \citet{w14}. Note that in some case the
  central X-mode or the O-mode in out-wings may be weak and only one
  mode is observable. The asymmetric intensity distribution of X-mode
  and O-mode emission caused by the additional co-rotation of
  particles can be used to understand the high degrees of linear
  polarization.
\end{enumerate}

In this paper, we studied the emission from curvature radiations with
or without the rotation involved, and the initial propagation (i.e.
``adiabatic walking'' and refraction) of the polarized waves within
pulsar magnetosphere. When the waves propagate outwards, other
propagation effects, e.g. wave mode coupling, should be considered
since it could alter the polarization states. For example,
$\Gamma_{\rm ad}$ decreases as the wave propagates outside until
$\Gamma_{\rm ad}\ll 1$. Near the region $\Gamma_{\rm ad} \sim 1$, the
wave mode coupling happens, which will lead the generation of circular
polarization \citep{wlh10}. As was demonstrated in \citet{wlh10} and
\citet{bp12}, the sense of circular polarization is closely related to
the polarization modes and PA curve gradient. For the same PA curve
gradient for a pulsar, the circular polarization generated by the wave
mode coupling should have opposite signs for the X-mode and O-mode in
different parts of profiles, which means the sign reversal of circular
polarization accompanying the orthogonal mode jump.

Our calculations can explain some observation facts. However, various
observation features exist, for example, the partial cone polarization
features \citep{lm88}, the evolvement of the polarization intensity
across frequencies \citep{vx97,yh06}, high degree of circular
polarization \citep{blh+88}, the polarization evolution for the
precession pulsar J1141-6545 \citep{mks+10}, etc, which need further
investigations. In our model, the plasma within the pulsar
magnetosphere is simply assumed to be cold, symmetric for electron and
positron and with a density distribution of uniformity, cone, core and
patches. However, the actual energy and density distributions of the
particles are not known. The plasma generated through the cascade
process in the polar cap region can be hot and has a wide energy
distribution \citep{ml10}. Meanwhile, the sparking process may also
lead to irregular density patches. Nevertheless, only if the O-mode
components are coupled to the superluminous O-mode can they leave the
magnetosphere, which suggests the intrinsical nonlinear process
\citep{ba86}. However, little is known about the nonlinear process up
to now. In our calculations for the coherent curvature radiation, the
coherency is simply treated by assuming the coherent bunch to be a
huge point charge. However, the actual coherent manner, both in the
bunching form of particles and the maser-type instabilities, will lead
to diverse radiation patterns. In our study, the dipole magnetosphere
is used to investigate the curvature radiation and propagation
processes, while the distortion of the dipole magnetosphere caused
either by pulsar rotation or by the polar cap current
\citep{dh04,kg12,kg13} is not considered. These factors will make the
final emergent radiation much more complicated \citep{bp12} and should
be investigated in the future.

\section*{Acknowledgments}

The authors thank the anonymous referee for helpful comments. This
work has been supported by the National Natural Science Foundation of
China (11273029 and 11003023) and the Young Researcher Grant of
National Astronomical Observatories, Chinese Academy of Sciences.

\bibliographystyle{mn2e}

\begin{thebibliography}{}

\bibitem[\protect\citeauthoryear{{Arons} \& {Barnard}}{{Arons} \&
  {Barnard}}{1986}]{ab86}
{Arons} J.,  {Barnard} J.~J.,  1986, ApJ, 302, 120

\bibitem[\protect\citeauthoryear{{Barnard} \& {Arons}}{{Barnard} \&
  {Arons}}{1986}]{ba86}
{Barnard} J.~J.,  {Arons} J.,  1986, ApJ, 302, 138

\bibitem[\protect\citeauthoryear{{Benford} \& {Buschauer}}{{Benford} \&
  {Buschauer}}{1977}]{bb77}
{Benford} G.,  {Buschauer} R.,  1977, MNRAS, 179, 189

\bibitem[\protect\citeauthoryear{{Goldreich} \& {Julian}}{{Goldreich} \&
  {Julian}}{1969}]{gj69}
{Goldreich} P.,  {Julian} W.~H.,  1969, ApJ, 157, 869

\bibitem[\protect\citeauthoryear{{Beskin}, {Gurevich} \& {Istomin}}{{Beskin}
  et~al.}{1988}]{bgi88}
{Beskin} V.~S.,  {Gurevich} A.~V.,    {Istomin} I.~N.,  1988, Ap\&SS, 146, 205

\bibitem[\protect\citeauthoryear{{Beskin}, {Gurevich} \& {Istomin}}{{Beskin}
  et~al.}{1993}]{bgi93}
{Beskin} V.~S.,  {Gurevich} A.~V.,    {Istomin} Y.~N.,  1993, {Physics of the pulsar magnetosphere}, Cambridge University Press, Cambridge 

\bibitem[\protect\citeauthoryear{{Beskin}, {Istomin} \& {Philippov}}{{Beskin}
  et~al.}{2013}]{bip13}
{Beskin} V.~S.,  {Istomin} I.~N., {Philippov} A.~A., 2013, Physics Uspekhi, 56, 164

\bibitem[\protect\citeauthoryear{{Beskin} \& {Philippov}}{{Beskin} \&
  {Philippov}}{2011}]{bp11}
{Beskin} V.~S.,  {Philippov} A.~A.,  2011, submitted to MNRAS, arXiv:1101.5733

\bibitem[\protect\citeauthoryear{{Beskin} \& {Philippov}}{{Beskin} \&
  {Philippov}}{2012}]{bp12}
{Beskin} V.~S.,  {Philippov} A.~A.,  2012, MNRAS, 425, 814

\bibitem[\protect\citeauthoryear{{Biggs}, {Lyne}, {Hamilton}, {McCulloch} \&
  {Manchester}}{{Biggs} et~al.}{1988}]{blh+88}
{Biggs} J.~D.,  {Lyne} A.~G.,  {Hamilton} P.~A.,  {McCulloch} P.~M.,
  {Manchester} R.~N.,  1988, MNRAS, 235, 255

\bibitem[\protect\citeauthoryear{{Blaskiewicz}, {Cordes} \&
  {Wasserman}}{{Blaskiewicz} et~al.}{1991}]{bcw91}
{Blaskiewicz} M.,  {Cordes} J.~M.,    {Wasserman} I.,  1991, ApJ, 370, 643

\bibitem[\protect\citeauthoryear{{Buschauer} \& {Benford}}{{Buschauer} \&
  {Benford}}{1976}]{bb76}
{Buschauer} R.,  {Benford} G.,  1976, MNRAS, 177, 109

\bibitem[\protect\citeauthoryear{{Cheng} \& {Ruderman}}{{Cheng} \&
  {Ruderman}}{1979}]{cr79}
{Cheng} A.~F.,  {Ruderman} M.~A.,  1979, ApJ, 229, 348

\bibitem[\protect\citeauthoryear{{Dyks} \& {Harding}}{{Dyks} \&
  {Harding}}{2004}]{dh04}
{Dyks} J.,  {Harding} A.~K.,  2004, ApJ, 614, 869

\bibitem[\protect\citeauthoryear{{Gangadhara}}{{Gangadhara}}{2010}]{gan10}
{Gangadhara} R.~T.,  2010, ApJ, 710, 29

\bibitem[\protect\citeauthoryear{{Gil} \& {Snakowski}}{{Gil} \&
  {Snakowski}}{1990}]{gs90}
{Gil} J.~A.,  {Snakowski} J.~K.,  1990, A\&A, 234, 237

\bibitem[\protect\citeauthoryear{{Han}, {Demorest}, {van Straten} \&
  {Lyne}}{{Han} et~al.}{2009}]{han09}
{Han} J.~L.,  {Demorest} P.~B.,  {van Straten} W.,    {Lyne} A.~G.,  2009,
  ApJS, 181, 557

\bibitem[\protect\citeauthoryear{{Han}, {Manchester}, {Xu} \& {Qiao}}{{Han}
  et~al.}{1998}]{hmx+98}
{Han} J.~L.,  {Manchester} R.~N.,  {Xu} R.~X.,    {Qiao} G.~J.,  1998, MNRAS,
  300, 373

\bibitem[\protect\citeauthoryear{{Jackson}}{{Jackson}}{1975}]{jac75}
{Jackson} J.~D.,  1975, {Classical electrodynamics}, Wiley, New York

\bibitem[\protect\citeauthoryear{{Kazbegi}, {Machabeli} \& {Melikidze}}{{Kazbegi} et ~al.}{1991}]{kmm91}
{Kazbegi} A.~Z.,  {Machabeli} G.~Z., {Melikidze} G.~I.,  1991, MNRAS, 253, 377

\bibitem[\protect\citeauthoryear{{Kumar} \& {Gangadhara}}{{Kumar} \&
  {Gangadhara}}{2012}]{kg12}
{Kumar} D.,  {Gangadhara} R.~T.,  2012, ApJ, 746, 157

\bibitem[\protect\citeauthoryear{{Kumar} \& {Gangadhara}}{{Kumar} \&
  {Gangadhara}}{2013}]{kg13}
{Kumar} D.,  {Gangadhara} R.~T.,  2013, ApJ, 769, 104

\bibitem[\protect\citeauthoryear{{Luo}, {Melrose} \& {Machabeli}}{{Luo}
    et ~al.}{1994}]{lmm94} {Luo} Q., {Melrose} D.~B., {Machabeli}
  G.~Z., 1994, MNRAS, 268, 159

\bibitem[\protect\citeauthoryear{{Luo} \& {Melrose}}{{Luo} \&
  {Melrose}}{2001}]{lm01}
{Luo} Q.,  {Melrose} D.~B.,  2001, MNRAS, 325, 187

\bibitem[\protect\citeauthoryear{{Lyne} \& {Manchester}}{{Lyne} \&
  {Manchester}}{1988}]{lm88}
{Lyne} A.~G.,  {Manchester} R.~N.,  1988, MNRAS, 234, 477

\bibitem[\protect\citeauthoryear{{Lyubarskii} \& {Petrova}}{{Lyubarskii} \&
  {Petrova}}{1998}]{lp98}
{Lyubarskii} Y.~E.,  {Petrova} S.~A.,  1998, A\&A, 333, 181

\bibitem[\protect\citeauthoryear{{Manchester}, {Kramer}, {Stairs}, {Burgay},
  {Camilo}, {Hobbs}, {Lorimer}, {Lyne}, {McLaughlin}, {McPhee}, {Possenti},
  {Reynolds} \& {van Straten}}{{Manchester} et~al.}{2010}]{mks+10}
{Manchester} R.~N.,  {Kramer} M.,  {Stairs} I.~H.,  {Burgay} M.,  {Camilo} F.,
  {Hobbs} G.~B.,  {Lorimer} D.~R.,  {Lyne} A.~G.,  {McLaughlin} M.~A.,
  {McPhee} C.~A.,  {Possenti} A.,  {Reynolds} J.~E.,    {van Straten} W.,
  2010, ApJ, 710, 1694

\bibitem[\protect\citeauthoryear{{McKinnon} \& {Stinebring}}{{McKinnon} \&
  {Stinebring}}{2000}]{ms00}
{McKinnon} M.~M.,  {Stinebring} D.~R.,  2000, ApJ, 529, 435

\bibitem[\protect\citeauthoryear{{Medin} \& {Lai}}{{Medin} \&
  {Lai}}{2010}]{ml10}
{Medin} Z.,  {Lai} D.,  2010, MNRAS, 406, 1379

\bibitem[\protect\citeauthoryear{{Melrose} \& {Stoneham}}{{Melrose} \&
  {Stoneham}}{1977}]{ms77}
{Melrose} D.~B.,  {Stoneham} R.~J.,  1977, Proceedings of the Astronomical
  Society of Australia, 3, 120

\bibitem[\protect\citeauthoryear{{Ochelkov} \& {Usov}}{{Ochelkov} \&
  {Usov}}{1980}]{ou80}
{Ochelkov} I.~P.,  {Usov} V.~V.,  1980, Ap\&SS, 69, 439

\bibitem[\protect\citeauthoryear{{Petrova}}{{Petrova}}{2006}]{pet06}
{Petrova} S.~A.,  2006, MNRAS, 366, 1539

\bibitem[\protect\citeauthoryear{{Radhakrishnan} \& {Cooke}}{{Radhakrishnan} \&
  {Cooke}}{1969}]{rc69}
{Radhakrishnan} V.,  {Cooke} D.~J.,  1969, Astrophys.Lett., 3, 225

\bibitem[\protect\citeauthoryear{{Radhakrishnan} \& {Rankin}}{{Radhakrishnan}
  \& {Rankin}}{1990}]{rr90}
{Radhakrishnan} V.,  {Rankin} J.~M.,  1990, ApJ, 352, 258

\bibitem[\protect\citeauthoryear{{Rankin} \& {Ramachandran}}{{Rankin} \&
  {Ramachandran}}{2003}]{rr03}
{Rankin} J.~M.,  {Ramachandran} R.,  2003, ApJ, 590, 411

\bibitem[\protect\citeauthoryear{{Ruderman} \& {Sutherland}}{{Ruderman} \&
  {Sutherland}}{1975}]{rs75}
{Ruderman} M.~A.,  {Sutherland} P.~G.,  1975, ApJ, 196, 51

\bibitem[\protect\citeauthoryear{{Stinebring}, {Cordes}, {Rankin}, {Weisberg}
  \& {Boriakoff}}{{Stinebring} et~al.}{1984}]{scr+84}
{Stinebring} D.~R.,  {Cordes} J.~M.,  {Rankin} J.~M.,  {Weisberg} J.~M.,
  {Boriakoff} V.,  1984, ApJS, 55, 247

\bibitem[\protect\citeauthoryear{{von Hoensbroech} \& {Xilouris}}{{von
  Hoensbroech} \& {Xilouris}}{1997}]{vx97}
{von Hoensbroech} A.,  {Xilouris} K.~M.,  1997, A\&AS, 126, 121

\bibitem[\protect\citeauthoryear{{Wang}}{{Wang}}{2014}]{w14}
{Wang} C., 2014, ApJ, submitted

\bibitem[\protect\citeauthoryear{{Wang} \& {Lai}}{{Wang} \& {Lai}}{2007}]{wl07}
{Wang} C.,  {Lai} D.,  2007, MNRAS, 377, 1095

\bibitem[\protect\citeauthoryear{{Wang}, {Lai} \& {Han}}{{Wang}
  et~al.}{2010}]{wlh10}
{Wang} C.,  {Lai} D.,    {Han} J.~L.,  2010, MNRAS, 403, 569

\bibitem[\protect\citeauthoryear{{Wang}, {Wang} \& {Han}}{{Wang}
  et~al.}{2012}]{wwh12}
{Wang} P.~F.,  {Wang} C.,    {Han} J.~L.,  2012, MNRAS, 423, 2464

\bibitem[\protect\citeauthoryear{{Wu}, {Manchester} \& {Lyne}}{{Wu}
  et~al.}{1993}]{wml93}
{Wu} X.~J.,  {Manchester} R.~N.,    {Lyne} A.~G.,  1993, MNRAS, 261, 630

\bibitem[\protect\citeauthoryear{{Xu}, {Liu}, {Han} \& {Qiao}}{{Xu}
  et~al.}{2000}]{xlh+00}
{Xu} R.~X.,  {Liu} J.~F.,  {Han} J.~L.,    {Qiao} G.~J.,  2000, ApJ, 535, 354

\bibitem[\protect\citeauthoryear{{You} \& {Han}}{{You} \& {Han}}{2006}]{yh06}
{You} X.~P.,  {Han} J.~L.,  2006, Chin.J.Astron.Astrophys., 6, 237

\end{thebibliography}

\label{lastpage}

\end{document}